\documentclass[proceedings, preprint]{rmaa}

\usepackage{paralist}

\usepackage{psfrag,color}

\usepackage{natbib}

\SetYear{2023}
\SetConfTitle{Prospects for low-frequency radio astronomy in S. A.}

\title{LLAMA Millimeter and Submillimeter Observatory. Update on its Science Opportunities.} 

\author{
  M. Fern\'andez-L\'opez,\altaffilmark{1} 
  P. Benaglia,\altaffilmark{1}
  S. Cichowolski,\altaffilmark{2}
  F. S. Correra,\altaffilmark{3}
  G. Cristiani,\altaffilmark{2}
  T.  P. Dominici,\altaffilmark{4}
  N. Duronea,\altaffilmark{5,6}
  G. Gimenez de Castro,\altaffilmark{2,7}
  J. R. D. Lepine,\altaffilmark{8}
  I. F. Mirabel,\altaffilmark{2}
  J. P. Raulin,\altaffilmark{7}
  H. Salda\~{n}o,\altaffilmark{9,10}
  L. Suad,\altaffilmark{2} and
  C. Valotto\altaffilmark{11,12}
  }

\altaffiltext{1}{Instituto Argentino de Radioastronom\'\i a (CCT-La Plata, CONICET; CICPBA), C.C. No. 5, 1894, Villa Elisa, Buenos Aires, Argentina (manferna@gmail.com).}
\altaffiltext{2}{Instituto de Astronom\'\i a y F\'\i sica del Espacio, CONICET–UBA, Argentina.}
\altaffiltext{3}{Microelectronics Laboratory of Polytechnic School, University of S\~{a}o Paulo, SP, P.O. Box 61548, CEP 05424-970, Brazil.}
\altaffiltext{4}{Instituto Nacional de Pesquisas Espaciais, 12227-010 S\~{a}o Jos\'e dos Campos, S\~{a}o Paulo, Brazil.}
\altaffiltext{5}{Instituto de Astrof\'\i sica de La Plata (UNLP-CONICET), La Plata, Argentina.}
\altaffiltext{6}{Faculdad de Ciencias Astron\'omicas y Geof\'\i sicas, Universidad Nacional de La Plata, Paseo del Bosque s/n, 1900, La Plata, Argentina.}
\altaffiltext{7}{Centro de R\'adio Astronomia e Astrof\'\i sica Mackenzie, Escola de Engenharia, Universidade Presbiteriana Mackenzie, Pr\'edio 45 T, cobertura, R. da Consola\c{c}\~{a}o, 896–7o andar–Consola\c{c}\~{a}o, S\~{a}o Paulo, Brazil.}
\altaffiltext{8}{Instituto de Astronomia, Geof\'\i sica e Ci\^{e}ncias Atmosf\'ericas; Universidade de S\~{a}o Paulo, Rua do Mat\~{a}o, 1226, S\~{a}o Paulo, SP, Brazil.}
\altaffiltext{9}{Instituto de Investigaciones en Energ\'\i a no Convencional, Universidad Nacional de Salta, C.P.: 4400, Salta, Argentina.}
\altaffiltext{10}{Consejo Nacional de Investigaciones Científicas y Técnicas (CONICET), Godoy Cruz 2290, CABA, CPC 1425FQB, Argentina.}
\altaffiltext{11}{Instituto de Astronom\'\i a Te\'orica y Experimental, CONICET-UNC, C\'ordoba, Argentina.}
\altaffiltext{12}{Observatorio Astron\'omico de C\'ordoba, Universidad Nacional de C\'ordoba, Laprida 854, C\'ordoba X5000BGR, Argentina.}

\shortauthor{Fern\'andez-L\'opez et al.}
\shorttitle{Science with LLAMA}

\listofauthors{M. Fern\'andez-L\'opez,  P. Benaglia, S. Cichowolski, F. S. Correra, G. Cristiani,  T.  P. Dominici, N. Duronea,   G. Gim\'enez de Castro,  J. R. D. L\'epine, I. F. Mirabel,  J. P. Raulin, H. Salda\~{n}o, L. Suad, \& C. Valotto}
\indexauthor{Fern\'andez-L\'opez, M.}
\indexauthor{Benaglia, P.}
\indexauthor{Cichowolski, S.} 
\indexauthor{Correra, F. S.}
\indexauthor{Cristiani, G.}
\indexauthor{Dominici, T. P.}
\indexauthor{Duronea,  N.}
\indexauthor{Gim\'enez de Castro, G.}  
\indexauthor{L\'epine, J. R. D.}
\indexauthor{Mirabel, I. F.}
\indexauthor{Raulin, J. P.}
\indexauthor{Salda\~{n}o, H.}
\indexauthor{Suad, L.}
\indexauthor{Valotto, C.}

\abstract{ The Large Latin American Millimeter Array (LLAMA for short) is a joint  scientific and technological undertaking of Argentina and Brazil whose goal is to install and to operate an observing facility capable of performing observations of the Universe at millimeter and sub-millimeter wavelengths. It will consist of a 12~m ALMA-like antenna with the addition of two Nasmyth cabins. LLAMA is located at 4850\,m above sea level in the Puna Salte\~{n}a, in the northwest region of Argentina. When completed, LLAMA will be equipped with
six ALMA receivers covering Bands 1, 2+3, 5, 6, 7, and 9, which will populate the two Nasmyth cabins. We summarize here the main ideas related with the Science that LLAMA could accomplish on different astronomical topics, gathered from the experince of a group of international experts on each field.   
}

\resumen{El proyecto Large Latin American Millimeter Array (LLAMA por sus siglas en ingl\'es) es un emprendimiento cient\'ifico y tecnol\'ogico de Argentina y Brasil, que tiene por objeto instalar y operar una facilidad observacional capaz de realizar observaciones del Universo en longitudes de onda milim\'etricas y submilim\'etricas. Consistir\'a de una antena de 12~m de di\'ametro tipo ALMA pero con la adici\'on de dos cabinas Nasmyth. LLAMA se ubicar\'a a 4850\,m de altura sobre el nivel del mar, en la Puna Salte\~{n}a, en el noroeste de la Rep\'ublica Argentina. Cuando sea completado, LLAMA poseer\'a seis receptores tipo ALMA que cubrir\'an las bandas 1, 2+3, 5, 6, 7 y 9, y que se distribuir\'an entre las dos cabinas Nasmyth. En esta contribuci\'on resumimos las principales ideas relacionadas con la Ciencia que LLAMA podr\'ia tratar de atacar en diferentes campos de la Astronom\'ia. Estos ideas fueron recabadas a partir de una reuni\'on de expertos internacionales en cada uno de estos campos.}

\addkeyword{Millimeter and Submillimeter Astronomy}
\addkeyword{LLAMA Telescope}

\begin{document}
\maketitle

\section{Introduction}
\label{sec:intro}
The Large Latin American Millimeter Array\footnote{https://www.llamaobservatory.org/} (LLAMA for short) is a joint  scientific and technological undertaking of Argentina and Brazil whose goal is to install and to operate an observing facility capable of performing observations of the Universe at millimeter and sub-millimeter wavelengths. It will consist of a 12-m ALMA-like antenna with the addition of two Nasmyth cabins. LLAMA is located at 4850 m above sea level in the Puna Salteña, in the northwest region of Argentina. When completed, LLAMA will be equipped with:
\begin{itemize}
\item Six ALMA receivers covering Bands 1, 2+3, 5, 6, 7, and 9, which will populate the two Nasmyth cabins.
\item Invited instruments, that be installed at the Cassegrain cabin.
\item The capability for carrying out simultaneous observations at two or more Bands.
\item The capability of carrying out continuum, spectral, polarization, and Solar observations.
\end{itemize}

The construction of the LLAMA observational facilities is underway and it is expected to reach the astronomical First Light by mid-2025. The antenna and its components are already at the high site and the antenna platform and anchor ring are already in their final place; the primary parabolic reflector is also assembled and temporal buildings are mounted for storage and rest; a building dedicated to the operations and maintenance is located in the nearby village of San Antonio de los Cobres; the recent incorporation of the Argentinean company INVAP has driven more organization to the project; and in 2022, a reorganized bi-national \lq\lq Steering and Science Committee\rq\rq, hosted an international workshop to revamp the project from the scientific point of view. 

During the week of the \textit{Science with LLAMA Workshop\footnote{https://www.llamaobservatory.org/ws2022/home.html}} (held in Salta City, Argentina, in September 2022) there were a series of talks and discussions led by experts on different astronomical topics such as: Planetary Atmospheres, Extragalactic Astronomy, LLAMA as part of a Very Long Baseline Interferometer (VLBI), Time Domain Astronomy, Solar Physics, Astrochemistry, Interstellar Medium, Magnetic Fields, and studies of Non-thermal sources like Supernova Remnants or Blazars. This document presents a brief compilation of the ideas that the speakers highlighted as feasible experiments that can be done with LLAMA. The content of the present document also includes portions of some of the abstracts and slides of those talks, along with specific and direct contributions from a few authors. The paper starts with a short overview of the science opportunities for LLAMA and it is organized in Sections that summarize the main ideas of the presentations and discussions of the 2022 Workshop aforementioned. 

\section{Key Science and Niches for LLAMA}
\label{sec:niches}
Submillimeter wavelength astronomy has made great progress in the last two decades, especially with the advent of interferometers like the Submillimeter Array (SMA), Northern Extended Millimeter Array (NOEMA), and the Atacama Large Millimeter/Submillimeter Array (ALMA). These instruments have brought great discoveries on the fine structures of a variety of astrophysical phenomena including star and planet formation, external galaxies, and the early universe. The latest Event Horizon Telescope (EHT) results on the shadow of supermassive black holes came from the connection of these interferometers to large single dish millimeter/submillimeter telescopes to form a global VLBI telescope. LLAMA will participate also in the next-generation Event Horizon Telescope (ngEHT) experiments. Its participation has been largely expected by the international community. LLAMA will also be an enormous asset for studying large scale structures (by including a new generation wide-field camera), magnetic fields, or surveying for rare phenomena on large fields of the sky. LLAMA will obtain images of molecular clouds in our own galaxy or entire maps of nearby galaxies. The ability to have a large amount of observing time in the submillimeter wavelengths will also be important for studying transient phenomena while having a big frequency bandwidth coverage will allow us to explore the chemical content and the spectral energy distributions (SEDs) of a diversity of sources, from the birth to the death of stars in the Milky Way and in other galaxies.

\section{Planetary Atmospheres}
\label{sec:planets}
Millimeter spectroscopy is a fantastic tool to expand our understanding of fundamental questions regarding the formation and evolution of the Solar System. Spectral signatures observed at the very high spectral resolution offered by heterodyne spectroscopy can reveal both the composition and dynamics of the probed atmospheres. Some of the open questions still unanswered in this field are: (i) how did the Solar System and the planets form? (ii) how do the planets work? (iii) how will the Solar System evolve? The study of the Giant Planets (GP) is crucial since they are the possible “architects” of the planetary systems. To know more about them, we must measure their composition and thermal structure in 3D and as a function of time, since these planets undergo seasonal changes and can be perturbed by the supply of exogenic material from comets and other bodies. Such measurements are key to constrain formation processes, chemistry, and atmospheric dynamics.

LLAMA will be able to detect and monitor the chemistry of past  comet impacts in GPs \citep[such as the Shoemaker-Levy~9 in Jupiter,][]{Lellouch1995,Bjoraker1996,Moreno2003,Moreno2006}, and survey any newly formed species.

LLAMA will allow us to directly measure the winds and temperatures in telluric planets such as Mars and Venus in millimeter wavelengths \citep[see e.g. experiments by][]{Shah1991,Lellouch1991}. To this end, a very high spectral resolution (of less or about 50~ms$^{-1}$) is needed.

The advice for LLAMA is to dedicate Bands 1-7 for time monitoring observations of: 

\begin{itemize}
\item Venus and Mars atmospheres. 
\item As a support to space missions (Mars orbiters and landers are not sensitive to the mesosphere which is probed by millimeter waves), as long as six new missions to Venus are expected for the next ten years.
\item Comet activity and composition along their orbital paths. These are key to constrain their primordial composition that may change because of thermal processing.
\end{itemize}

LLAMA may try new detections e.g., in Band 9:
\begin{itemize}
\item Probe the deep composition of Giant Planets using large-band spectroscopy through tropospheric species which produce absorption lines (e.g., PH$_3$ is still undetected in Uranus and Neptune).
\end{itemize}

\section{LLAMA as part of VLBI Networks}
\label{sec:vlbi}
LLAMA is interested in forming part of Very Long Baseline Interferometry networks. The project is specially interested in entering the next generation Event Horizon Telescope (ngEHT) collaboration and it is seen as a key station for that experiment. The ngEHT project will quadruple bandwidths, increase observing frequency, and double the number of dishes in the global VLBI array. One of the most notable differences between the EHT and ngEHT, is the inclusion of new telescopes at new geographical sites to dramatically improve the range of interferometric baselines.  For this array, LLAMA can play an important role by replacing the baselines with APEX and ALMA, and/or providing a high signal-to-noise ratio (S/R) short baseline with ALMA (in the milliarcsec scales), aiding to fill up the uv-plane in the $\sim$200~km range, and providing redundancy with baselines using ALMA and APEX, which may provide advantages for phase closure calibration, absolute amplitude calibration, and polarization.

Regarding the science with the ngEHT, the endeavor of this project is mainly to make movies of the Sgr~A$^*$ massive black hole in the Galactic Center. Since the time variability scales of the black hole in Sgr~A$^*$ is expected to be from the 10s second scale to the 1 hour scale, to frame the actual structure, the ngEHT will try to make a big jump in sensitivity upgrading the backend and the receivers of the antennas, and adding more elements into the array. These enhancements will allow the ngEHT to capture the dynamics of black holes through real-time and time-lapse video. 

Also, the improvements will be used to: (i) image and make a time-lapse movie of the jet launch in M87, which may address questions related to the jet powering mechanism (gravitational vs spin energy), and the physical conditions in the base of the jet; (ii) make black hole movies of Sgr~A$^*$ to analyze the horizon scale flares and the magnetic field dynamics. These can also be used to test General Relativity to a better precision by resolving the photon ring structure, specifically focusing on the $n=1$ ring. The ngEHT collaboration has 8 science working groups, each studying a different topic: Fundamental Physics, Black holes and their Cosmic context,  Jet launching,  Accretion,  Transients,  New horizons, Algorithms and inference, and History, Philosophy, and Culture. 

Beyond the black hole studies in Sgr~A$^*$ and M87, the ngEHT has been proposed to play a role in other kind of studies such as: (i) probing Active Galactic Nuclei (AGN) jets from pc to kpc scales toward certain blazars, quasars, and radio galaxies such as Centaurus A, 3C279 or 3C273; (ii) analyzing maser spectral lines in Galactic star-forming regions and evolved stars; and (iii) studying the polarization and magnetic fields in Sgr~A$^*$, M87 and a larger sample of AGNs.

LLAMA also envisions being part of the Global 3-mm VLBI Array (GMVA), which corresponds to Band~3. This would be an important step towards expanding the network coverage in the Southern Hemisphere. Another challenge that LLAMA could undertake is to perform the first interferometric observation in Band~9. This could be done along with APEX or ALMA, for example, and would be a paramount success for the project. LLAMA will work with a best effort basis in getting a VLBI system ready to enter the EHT or ngEHT as soon as possible.  

\section{Time Domain (sub)mm Astronomy}
\label{sec:timedomain}
All the objects in the Universe change with time. From the flares in the solar surface to the jets associated with supermassive black holes at the center of galaxies, including the variability detected in several maser species in different astronomical scenarios. Time domain submillimetre observing is still in its infancy, and it is a window of opportunities for future facilities like LLAMA, along with the importance of ever-improving submillimetre calibration techniques. However, it is difficult and “astronomically expensive” to monitor different objects for long times; not to mention doing it at high frequencies, which needs excellent weather conditions and are specially challenging from an operational point of view. Another approach for LLAMA programs is to team up with smaller telescopes like CCAT, which will be monitoring large portions of the southern sky daily, to make dedicated follow-ups of intriguing or special cases. From a multi-wavelength and multi-messenger astronomical perspective, LLAMA will have synergy with new important observatories that are under construction, namely the Rubin Observatory, through the \lq\lq Legacy Survey of Space and Time\rq\rq, and the Cherenkov Telescope Array (CTA).

Long-term projects carried out with the James Clerk Maxwell Telescope (JCMT), can be taken as good examples of time-domain studies of different objects at millimeter wavelengths \citep[e.g.,][]{Herczeg2017,Lee2021}. For instance, the 7-years study  toward nearby molecular clouds where stars form has found that about 1/3 of the embedded protostars show continuum flux variations with time. This kind of studies have shown that even with a coarse angular resolution the analysis of multi-wavelength time monitoring data can provide information about the innermost structures and accretion events of protostellar disks. Hence, time monitoring can provide relevant clues about the physical properties of these objects. In the study of young stellar objects, the timescale variations can be related to different scale sizes depending on the processes involved \citep[e.g.,][]{Johnstone2013,Fischer2023}. In the work produced using the JCMT protostellar monitoring, there are different types of variations. There are quasi periodic variations with periods ranging from 2-6 years, but there are also truly transient, single-event detections of phenomena that need dedicated supplementary observations for a correct understanding \citep[e.g., the magnetic reconnection outburst event shown in][]{Mairs2019}.

In addition to the intrinsic scientific value of these works, time domain experiments can bring a beneficial side effect for the telescopes in terms of calibration strategies and systematics \citep[e.g., pointing, telescope scheduling, relative flux calibration between different epochs, etc;][]{Mairs2017,Francis2020}.

During the 2022 Workshop many other transient phenomena were identified as plausible interesting targets for the LLAMA observatory. The time scale variations range between a few seconds (or even subsecond scales) to several years. We summarize in a list here some of the most  relevant objects, including a few comments for some of them.
\begin{itemize}
\item Microquasar activity.  There are about 25 Galactic and 5 extragalactic microquasars identified to date\footnote{\url{www.aim.univ-paris7/CHATY/Microquasars/microquasars.html}}. These accreting binary systems produce relativistic jets and have not been largely observed at (sub)millimeter wavelengths. LLAMA can extend the sample of observed objects and trigger coordinated multi-band observations with observatories working at other wavelengths. This can be important to quantify the link between the innermost disk and the base of the jet \citep{Corbel2013}, to break SED model degeneracies, and to understand the origin of the variabilities and the flaring mechanisms.  LLAMA can also explore variations in the polarization fraction of the jet (SED and polarization monitoring). 
\item Planet winds. As commented before, LLAMA will allow us to monitor the winds on the atmospheres of Solar System planets.
\item Comets and comet impacts with Giant Planets. Monitoring the evolution of the chemical composition of comets, and the impact of comets on Giant Planets is another relevant experiment that LLAMA can carry out. 
\item Solar activity. LLAMA will be able to track flares, sunspots, 3-minute oscillations, to make synoptic maps, etc. Solar transients at (sub)millimeter wavelengths are part of the important scientific niches for LLAMA. 
\item Molecular line emission in evolved stars. This kind of monitoring can reveal details about the innermost structure and size of the circumstellar environment of evolved stars \citep[e.g.,][]{Pardo2018}. The observed maser variability is also included in this type of time monitoring \citep[e.g.,][]{Fonfria2018}.
\item Blazars and quasars. LLAMA will routinely monitor quasars as secondary flux calibrators, but the study of their light curves can also be an interesting matter for the analysis of flares or bursts \citep[e.g.,][]{Hovatta2015}. These brightest objects at radio wavelengths present jets, with components moving away from the core sometimes at superluminal velocities. These components are attributed to shocks propagating with relativistic velocities along jets that form small angles with the line of sight. When the shocks are generated, close to the central engine, the size of the emitting region is very small and optically thick at radio frequencies; the synchrotron emission is detected only at optical-soft X-ray frequencies, and the Inverse Compton emission at $\gamma$-rays. As the shocks propagate, the electrons lose energy first by Inverse Compton emission, afterwards by synchrotron emission and finally by adiabatic expansion. As the size increases, the source becomes optically thin and can be detected at millimeter  wavelengths. The delay between the start of variability at different wavelengths (which can be of several months) gives important information about the physical conditions of the shock and of the underlying jet. Radio variability monitoring of this kind of sources is nowadays restricted to a few radio telescopes, most of them in the northern hemisphere. LLAMA can make a great contribution to this field.
\item Fast Radio Burst. Accretion events of high-mass protostars. LLAMA can be used to monitor these objects in search of or giving support to detections of unique bursts such as the one experienced by NGC~6334~I \citep{Hunter2017}.
\item Maser monitoring at submillimeter wavelengths. LLAMA will allow us to easily monitor maser transitions, which may allow us to better understand their pumping mechanisms and the reasons for missing maser transitions in certain regions and at certain frequencies. The LLAMA contribution, especially at high frequencies, can be very important in this topic.   
\end{itemize}

\section{Solar Physics}
\label{sec:solar}
Solar observations in the (sub)millimeter range can measure the chromosphere, an upper layer of the solar atmosphere which has been poorly understood yet. The absolute brightness temperature distribution and its spatial and time variations can be measured in single-dish observations, which can be used to understand the atmospheric structures and possible heating mechanisms of the chromosphere. In addition to helping to understand the physical processes at work on the Sun, several of the solar investigations proposed through LLAMA will make a great synergy with studies of space weather.

LLAMA will expand the sensitivity and the frequency coverage of current observations carried out by the Solar Submillimeter Telescope (SST). Some of the niches for LLAMA are related to separate techniques.
For instance, the solar monitoring at different (sub)millimeter wavelengths of the continuum emission can provide clues about the vertical structure (physical conditions) of the chromosphere and the sunspots. LLAMA can even spatially resolve sunspots, with the advantage of single-dish over interferometric observations when trying to derive temperatures, since it will not be affected by the missing zero-spacing. In addition, time monitoring is expected to provide new detections of the so-called THz flares.  LLAMA can also contribute by measuring magnetic fields through polarization observations. Currently, there are no measurements by ALMA using circular polarization (Zeeman effect) observations, or polarization using lines (GK-effect). Regarding lines, observations of Radio Recombination Lines (RRLs) will be relevant for post-flare loops and off-limb studies. Finally, LLAMA’s multi-wavelength capability will play an important role in determining the acceleration mechanisms for relativistic particles in flares. 

LLAMA instrumental facility will bring new clues for Solar Physics open questions,
both during quiescent and active periods. The main (radio) emission mechanism at work during low activity periods is of thermal origin, called thermal free-free or bremsstrahlung emission. Its optical depth $T_{ff} \sim N_e^2 / T_e^{1.5}$ where $N_e$ is the electron density and $T_e$ the ambient temperature, with small departures below the transition region to account for H-opacity where the plasma is not fully ionized. Therefore, LLAMA multi-(high)frequency observations will provide important diagnostics to study the vertical structure of the cold and dense solar atmosphere layers above the quiet disk. This vertical structure is still unknown and different models exist although only a few studies tried to validate them against high-frequency radio observations \citep{delaLuz2011,Loukitcheva2014}. Validation of these models can also be tested by combining LLAMA and mid-infrared observations. The same can be said on the solar atmosphere above active regions, and above sunspots. 

Single-dish LLAMA observations will spatially resolve both the umbra and penumbra of sunspots in the low chromosphere, with a clear advantage from using (radio)interferometric observations only. Early works at mm wavelengths helped to understand the atmosphere above sunspots in the upper chromosphere \citep{Shimojo2017,Iwai2015}. LLAMA observations will also be relevant to understand other atmospheric structures, e.g. coronal plumes, and phenomena, e.g. 3-minutes oscillations, both important to better characterize the quiet Sun and its dynamics. On longer timescales, synoptic studies will also be possible using LLAMA, for example by performing a solar map per week. Such observations will bring new clues on the polar brightening phenomena, the local magnetic field, and how it evolves in time as a function of the solar activity cycle. One of the main unknown solar parameters is its magnetic field ($B$). Knowledge of $B$ is fundamental since we believe that the amount of energy released during solar flares is previously stored in the form of magnetic energy. The temperature gradient existing in the solar chromospheric plasma will allow LLAMA to bring information on the local magnetic field by providing polarization measurements. 

The principal phenomenon of the active Sun is the solar flare, which heats the ambient plasma to tens of MK, and accelerates particles, electrons, protons, and heavier nuclei to tens of MeV, few Gev, and few MeV per nucleus. Energetic electrons will radiate in the magneto-active plasma of the solar atmosphere in the form of synchrotron emission, at a frequency given by $\gamma^3$ $FB$ where $\gamma$ is the Lorentz factor of the particle and $FB$ is the local electron gyrofrequency. For this reason, high frequencie's LLAMA capability is very well suited to study the highest energy electrons accelerated during flares. By providing electron energy distribution and high-energy cutoff, LLAMA is also essential to better understand the acceleration mechanisms, which are poorly known. 

High-frequency radio spectra during flares are probably a composition of different emission mechanisms: optically thin (Ot) synchrotron from mildly-relativistic electrons, optically thick (OT) synchrotron from highly-relativistic electrons giving rise to the so-called THz spectrum \citep{Kaufmann2004}, OT bremsstrahlung from the dense and cold plasma in the chromosphere \citep{Trottet2015}, Ot bremsstrahlung from hotter plasma in the corona \citep{Raulin1999}, synchrotron from positrons \citep{Trottet2008}. All these mechanisms are incoherent in nature, but a coherent plasma mechanism, the Coherent Synchrotron Radiation (CSR) was also proposed \citep{Kaufmann2004}. LLAMA multi-frequency and polarization measurements, combined with other diagnostics provided by ALMA and the SST will therefore be able to get the precise shape of the high-frequency flare spectra, determine the nature of the radiation, estimate the counterpart of each of the above mechanisms, identify and separate electron and positron radiation. Finally, we remember that the above challenges critically depend on a precise radio flux determination which in turn is strongly affected by atmospheric attenuation. Therefore, special efforts are also needed to estimate how the incoming solar radiation is affected by the Earth's atmosphere at $\sim 5000$~m.

\section{Interstellar Medium}
\label{sec:ism}
The Interstellar Medium (ISM) may span a wider range of topics and physical phenomena than any other branch of astronomy. Already nearly a half-century ago, Spitzer's classical text includes topics ranging from molecular, atomic, and ionized clouds, to magnetic fields, dust grains, cosmic rays, and shocks and explosive motions of HII regions and supernova remnants. More modern treatments include topics such as star formation, stellar winds and feedback mechanisms, turbulence, and astrochemistry, among others. Many phenomena of the interstellar medium require higher angular resolution than a single-dish telescope can provide; this is especially true of research areas such as high-mass star formation. Nevertheless, by shrewd use of observational probes, and by careful selection of the phenomena to study, single-dish telescopes such as LLAMA have a great deal to offer to studies of the ISM.

\subsection{Star Formation Observations}
\label{sec:sfr}
Although LLAMA is not intended for high-angular resolution studies, there are some tricks that can overcome this situation. For instance, high-spectral resolution observations ($\sim0.1$ km~s$^{-1}$) of chemical thermometers, such as the CH$_3$CN K-ladder, can provide insight of the innermost layers of, for example, Hot Molecular Cores \citep[HMCs,][]{Rosero2013} and other objects of the ISM. 

However, LLAMA should exploit the science that does not need high-angular resolution, and the ISM is full of not well-known large-scale phenomena. For instance, LLAMA can contribute to studies about the Nitrogen fractionation of the local ISM by measuring the ratio between N-bearing isotope species \citep[e.g., HCCCN and HCCC$^{15}$N; see ][]{HilyBlant2018}. The differences in the abundance ratios of these isotopologues between solar system objects and the local ISM have raised a series of interesting issues (for instance, if the ratio is “inherited” or “processed”, and hence, the relationship between prestellar clouds, protostellar cores and planet formation). LLAMA will also play an important role in revealing the gas kinematics and coarse morphology of large-scale outflows from protostars and accretion streamers ($\sim0.1$~pc scales) toward low- and high-mass protostars \citep[e.g.,][]{Pineda2023}, and, at even larger scales, of expanding HII regions, or nearby filamentary molecular clouds, which will aid in digging into the formation and evolution of these structures \citep{Hacar2023}. In the same way, the triggering of Giant Molecular Cloud formation after the collision of supershell structures \citep{Inutsuka2015,Suad2022} can be probed kinematically with LLAMA \citep[e.g.,][]{Fujii2021} in cases previously selected by the analysis of specific cataloged HI structures \citep{Suad2014}. 
LLAMA can also start systematic studies of high-energy outflow tracers to state general properties of these phenomena, such as the mass, momentum or energy injected; with these, it will be possible to estimate the contribution of star-formation feedback effects into the energy budget of the ISM.

\subsection{Spectral Energy Distributions from Young Stars and Brown Dwarfs}
\label{sec:sed}
The capability of carrying out simultaneous multi-wavelength continuum observations of LLAMA can also make important contributions to the determination of Spectral Energy Distributions (SEDs) of disks from low-mass Young Stellar Objects (planet-forming and debris disks) and Brown Dwarfs (BDs), among other sources of the Interstellar Medium. For instance, surveys already carried out over hundreds of sources with ALMA \citep[e.g.,][]{Cieza2019}, can be complemented by observing them at the too demanding frequency Bands 5 and 9. Just a few observatories would be able to perform such kinds of observations successfully. Obtaining a significant sample of sources at these frequencies will probably constitute the most complete (sub)mm SED sample ever studied.

In the case of BDs, there is still no consensus about how they are formed \citep[e.g.,][]{Pinilla2013,Pinilla2017,Rilinger2021}. LLAMA can contribute to detect and confirm more of these sources (they have their SED peak in the submillimeter regime) by surveying close molecular clouds and computing the SED between Band~9 and other at a lower frequency. LLAMA can also measure and characterize their disks by deriving, for instance, the gas and dust masses, disk sizes, and $\alpha$-viscosity parameters from adding more constraints to their SEDs.

\subsection{Star Formation triggered by Microquasar Jets in the Milky Way}
\label{sec:uquasar}
Extragalactic observations had shown that jets launched by actively accreting supermassive black holes generate massive outflows that quench star formation in low-density gas clouds. However, by compressing high-density molecular clouds, black-hole outflows can, on the contrary, enhance star formation. Because of the expected high molecular gas densities in galaxies of the early universe, this star formation triggering mechanism should have been very important. In this context, the search and detailed follow-up observations at millimeter wavelengths of this star formation-triggering mechanism in the Milky Way \citep[e.g.,][]{Chaty2001}, may provide important information for studies on starburst galaxies at cosmological distances.

\subsection{Supernova Remnants and Cosmic Ray Acceleration}
\label{sec:snr}
Supernova remnants (SNRs) are also among the ISM objects whose study will be addressed with LLAMA. LLAMA will search for interactions between molecular clouds and supernova remnants to pinpoint places where particle acceleration (and the consequent cosmic ray production) could be taking place. Some open questions still remain in this field, such as if all kinds of supernovae accelerate particles, in which phase of their evolution and if the maximum energy of these particles can go beyond 1~PeV \citep{deAngelis2018}. Shock chemistry and kinematics studies are two avenues for LLAMA experiments, which will be complemented with high-energy observations, in order to learn about possible acceleration mechanisms. Observations of RRLs will also be carried out to measure the physical properties of the ionized gas produced at the SNR-molecular cloud interaction places.

\subsection{Pulsars and Magnetars}
\label{sec:pulsar}
Other ISM objects, such as pulsars or magnetars do not need high-angular resolution to be investigated. Actually, studying this kind of objects in the (sub)millimeter regime has been pioneered only since recently \citep[e.g.,][for the first detections of magnetars in millimeter and submillimeter bands, respectively]{Torne2015,Torne2022}, and will probably provide a new fruitful area of research for LLAMA. Also, the (sub)millimeter bands are best suited for searching for new magnetars and pulsars toward the Galactic Center and in nearby galaxies \citep{Torne2015}.

\subsection{Fundamental Physics}
\label{sec:physics}
Despite LLAMA's relatively low collective area and hence sensitivity, it is expected that Brazil and Argentina communities have a lot of time available for dedicated experiments. This will allow for deep observations and monitoring programs (see Section \ref{sec:timedomain} dedicated to Time Domain experiments). Using this, another sort of more exotic experiments that LLAMA could face by observing ISM objects is related to Fundamental Physics such as the ratio between the mass of the proton and the electron, which depends only on atomic constants. This ratio has been suggested to vary with local matter density, by means of the light scalar field coupling to matter \citep{Dapra2017}. Besides, extragalactic spectral line measurements over a range of redshift values could tell us if the fundamental constants vary with the age of the Universe \citep[e.g.,][]{Truppe2014}.

\section{Astrochemistry}
\label{sec:astrochem}
LLAMA is uniquely positioned to carry out studies on the water content of different astronomical objects, such as low- and high-mass star forming regions, the circumstellar environment of evolved stars, or the atmospheres of planets and moons. Although this kind of observations are constrained to excellent weather conditions at 183~GHz (Band~5), the investment could pay-off scientifically for sensitive observations and mapping of molecular clouds, and/or time monitoring of molecular clouds and evolved stars. 

LLAMA's time availability can be key as well to make extremely sensitive molecular line surveys in molecular clouds and evolved stars. LLAMA has to look for what other telescopes have not achieved in previous line surveys. Picking up a handful of sources and improving the previous sensitivities by a factor of ten, will be key for the success of these experiments. As an example, surveying a Dark Cloud with a low kinetic temperature (hence, very narrow line-widths) can avoid line confusion which may provide the detection of new molecular transitions. LLAMA will possibly allow for the detection and identification of new molecules in space \citep[e.g.,][]{Cernicharo2021}. For these new identifications, LLAMA's high frequency, multi-band availability, and a large bandwidth coverage are crucial. Bands 3, 5, 7, and 9 are probably the most unexplored (except for observations with Herschel/HIFI). Besides the observations, a coordinated effort with experimental astrophysics laboratories could optimize the identification of new molecular species.

LLAMA will also allow large-scale mapping of key molecular tracers in several rotational transitions. Using medium sensitivity, LLAMA can probe molecules such as HCN, HNC, CN, HCO$^+$, CS, HC$_3$N, etc. Using a multi-pixel camera, LLAMA will survey the continuum emission of large-scale portions of molecular clouds as well. LLAMA could also make extremely sensitive maps of all clumps found in the large-scale maps, with the goal of detecting isotopologues and vibrationally excited lines (study the physical properties and trace evolutionary paths). LLAMA can also be used as a pathfinder in high-angular resolution studies with ALMA by characterizing the physical properties (i.e., densities, temperatures through multi-line mapping) from the large to the smallest scales in Band~9. 

LLAMA will aid in searching for links between the chemistry and the multiplicity in the young stellar systems of nearby molecular clouds \citep[e.g.,][]{Murillo2023}. The chemical diversity at the scale of protostellar envelopes has not been yet thoroughly investigated and, along with studies correlating the physical properties of the envelopes and the multiplicity, will be an important scientific opportunity for LLAMA. 

A special object for Astrochemistry studies are the so-called Hot Molecular Cores (or HMC; massive young stellar objects in a warming stage where the ice mantle content is released into the gas phase surrounding the hot compact embryo of the future star). Their spectra are very profuse in molecular lines, presenting some of the richest chemical spectra observed at (sub)millimeter wavelengths. LLAMA will probe the content of Complex Organic Molecules, and through detailed analysis of the HMC's physical properties and the fractionation of isotopologues of certain molecules such as HC$_3$N \citep[e.g.,][]{Duronea2019}, will help to describe the chemical reactions that can be used as, for instance, chemical clocks. The evolutionary stage of different Interstellar Medium objects can also be probed by studying the deuteration of different species \citep[e.g.,][]{Fontani2015}.
 
Regarding maser species, LLAMA will probe different transitions and vibrational states simultaneously, which is crucial to find out the physical conditions and investigate the pumping mechanism producing the maser effect. For instance, just recently, \citet{Breen2019} have reported the detection of six new methanol maser transitions, three of which are torsionally excited. The latter transitions are a ``missing link'' in maser pumping models. Monitoring or blind searches, may be both viable options for the LLAMA observatory.

\section{Magnetic Fields}
\label{sec:bfield}
Magnetic fields are thought to play a key role in various stages of star formation over a large range of scales. Most of the recent works in this research area have shown the power of dust polarization observations for probing the role of magnetic fields in star formation and interstellar medium evolution, and the importance of polarimetry for constraining dust properties. However, despite the wealth of data published during the last few years, there are still many gaps in our knowledge about magnetic fields in the ISM that LLAMA can explore \citep[e.g.,][]{Pattle2022a}. For once, the effects that the star-formation feedback in its different manifestations (expanding HII regions, outflows from protostars, ionization fronts and photodissociation regions, etc), and also the effects from SNRs, produce in the magnetic field distribution of their surroundings, and the results that these disturbances trigger, are mostly unknown or not well settled. LLAMA will have the opportunity to follow-up the dust polarization surveys carried out with the JCMT toward nearby molecular clouds in the skies of the southern hemisphere \citep{Pattle2022a}.

Other experiments that LLAMA will carry out involving dust grain polarization are related to: 
different dust alignment mechanisms \citep{Andersson2015} and the grain magnetic alignment (the mainly accepted theory for dust grains alignment in the ISM) itself \citep[e.g.,][]{Pattle2019};
magnetic field distribution in outflow cavities \citep[e.g.,][]{Pattle2022b};
magnetic field distribution in nearby galaxies \citep[see e.g. M82,][]{Pattle2021,LopezRodriguez2022};
test dust formation/destruction in the envelopes of Asymptotic Giant Branch stars and/or the Supernova Remnants \citep[e.g.,][]{Chastenet2022}.
In addition to the open scientific problems regarding magnetic fields, polarization observations, from which the information is collected, are very challenging at (sub)millimeter wavelengths. Linear polarization observations of molecular lines at (sub)millimeter wavelengths (which may be the golden key to map the magnetic field distribution in high-density regions of the ISM) are still very scarce. Single-dish detections on sensitive molecules could be a pathfinder for dedicated interferometric experiments. Also, circular polarization molecular measurements toward high-density regions are still lacking to measure the Zeeman effect and therefore provide a direct measurement of the magnetic field's strengths. These have only been successful toward a few sources using the IRAM 30~m telescope \citep[e.g.,][]{Crutcher2012}, and thereby LLAMA has room for new discoveries here as well.

\section{Extragalactic Astronomy}
\label{sec:extragal}

\subsection{Mega Masers}
\label{sec:megamaser}
LLAMA would not need high-angular resolution to study the inner parts of Active Galactic Nuclei orbited by clouds showing water maser emission. For instance, in \citet{Haschick1994}, the authors could derive the presence of a rotating ensemble of masing clouds in radii of about 0.1~pc around the nucleus of a Seyfert~2 galaxy, by measuring the velocity drifts of different spectral features in 7-year monitoring toward NGC4258.

\subsection{Nearby Galaxies}
\label{sec:neargal}
The relatively short distance to the Magellanic Clouds makes these galaxies interesting targets to study the interstellar medium in low metallicity environments, where the HI gas dominate the total gas mass, the fraction of molecular and dust components are extremely low, and the radiation field is several times higher than that of the Milky Way \citep{Jameson2016,Jameson2018,DiTeodoro2019}. Such properties would be similar to those found in the early Universe. Therefore, studying these galaxies would help us to understand the star-formation at high spatial resolution in a younger Universe than the current time. LLAMA will be able to map large portions of the Magellanic Clouds in different transitions of the CO and other molecules like HCO$^+$, CS, or SiO, with a spatial resolution between $\sim$2-10 pc. LLAMA will allow us to study the physical conditions of low and high-density star-forming regions, determine global properties like the CO-to-H$_2$ conversion factor molecular abundance, and study the kinematic properties of the gas in low metallicity regions \citep{Bolatto2008,Wong2011,Saldano2023}. These observations will be fundamental to compare with those obtained toward very distant low metallicity galaxies (e.g., WLM, located at $\sim$985~kpc) observed by the ALMA telescope at a similar spatial resolution \citep{Rubio2015}, which can therefore help to constrain the main properties of their gas used in semi-analytic galaxy formation models.
 
In other galaxies, LLAMA will allow us to investigate the morphology and kinematics of the bars of nearby galaxies, thought to transfer angular momentum outwards, channel inflow of gas resulting in enhanced star formation at the nucleus, and fuel AGN. The pattern speed of bars is a crucial parameter to understanding the evolution of spiral galaxies.

\vspace{0.3cm}
{\bf Aknowledgements:} We acknowledge the contributions and ideas from all the assistants to the LLAMA's and IAR's Workshops that have been summarized in this work. We thank specially the eight invited speakers to the LLAMA Workshop: Drs. Thibault Cavalie,  Jos\'e Cernicharo, Shepherd Doeleman, Paul Ho, Kazumasa Iwai, Douglas Johnstone, Stanley Kurtz, and Kate Pattle. Many parts of abstracts, talks and the discussions have been literally extracted, with the agreement of the respective authors, and are part of the text included here.  A complete list of speakers, assistants, and committees related to the LLAMA's and IAR's Workshops can be found in their respective webpages: https://www.llamaobservatory.org/ws2022/ and https://congresos.unlp.edu.ar/iar60ws/. This work owes to everyone of them.



\end{document}